\documentclass[conference]{IEEEtran}

\IEEEoverridecommandlockouts
\usepackage{cite}
\usepackage{amsmath,amssymb,amsfonts}
\usepackage{algorithmic}
\usepackage{graphicx}
\usepackage{textcomp}
\usepackage{xcolor}
\usepackage{soul}

\def\BibTeX{{\rm B\kern-.05em{\sc i\kern-.025em b}\kern-.08em
    T\kern-.1667em\lower.7ex\hbox{E}\kern-.125emX}}
\begin{document}

\makeatletter
\newcommand{\linebreakand}{%
  \end{@IEEEauthorhalign}
  \hfill\mbox{}\par
  \mbox{}\hfill\begin{@IEEEauthorhalign}
}
\makeatother

\title{A Versatile Data Fabric for Advanced IoT-Based Remote Health Monitoring\\

}
\author
{\IEEEauthorblockN{
Italo Buleje\IEEEauthorrefmark{1},
Vince S. Siu\IEEEauthorrefmark{1},
Kuan Yu Hsieh\IEEEauthorrefmark{1}\IEEEauthorrefmark{5}\IEEEauthorrefmark{6},
Nigel Hinds\IEEEauthorrefmark{4},
Bing Dang\IEEEauthorrefmark{1},\\
Erhan Bilal\IEEEauthorrefmark{1},
Thanhnha Nguyen\IEEEauthorrefmark{2}\IEEEauthorrefmark{3}\IEEEauthorrefmark{7},
Ellen E. Lee\IEEEauthorrefmark{2}\IEEEauthorrefmark{3}\IEEEauthorrefmark{7},
Colin A. Depp\IEEEauthorrefmark{2}\IEEEauthorrefmark{3}\IEEEauthorrefmark{7}\IEEEauthorrefmark{8},
Jeffrey L. Rogers\IEEEauthorrefmark{1}}
\\
\IEEEauthorblockA{
\IEEEauthorrefmark{1}\textit{Digital Health, IBM T.J. Watson Research Center, Yorktown Heights, NY, USA}\\
\IEEEauthorrefmark{4}\textit{Emerging Technology Engineering, 
IBM T.J. Watson Research Center, Yorktown Heights, USA}\\
\IEEEauthorrefmark{5}\textit{Department of Electrical and Computer Engineering, College of Electrical and Computer Engineering,}\\ \textit{National Yang Ming Chiao Tung University, Hsinchu 30010, Taiwan}\\
\IEEEauthorrefmark{6}\textit{Institute of Biomedical Engineering, College of Electrical and Computer Engineering,}\\ \textit{ National Yang Ming Chiao Tung University, Hsinchu 30010, Taiwan}\\
\IEEEauthorrefmark{2}\textit{Department of Psychiatry, University of California San Diego, La Jolla, CA, USA}\\
\IEEEauthorrefmark{3}\textit{Sam and Rose Stein Institute for Research on Aging, University of California San Diego, La Jolla, CA, USA}\\
\IEEEauthorrefmark{7}\textit{USA IBM-UCSD Artificial Intelligence for Healthy Living Center, La Jolla, CA, USA}\\
\IEEEauthorrefmark{8}\textit{Veterans Affairs San Diego Healthcare System, San Diego, CA, USA}\\
Emails: [\{ibuleje, vssiu, nhinds, dangbing, jeffrogers\}@us.ibm.com, kuan.yu@ibm.com\\
\{t2vo, eel013, cdepp\}@health.ucsd.edu\\
}}

\maketitle
\IEEEpeerreviewmaketitle

\IEEEpubidadjcol
\begin{abstract}

This paper presents a data-centric and security-focused data fabric designed for digital health applications. With the increasing interest in digital health research, there has been a surge in the volume of Internet of Things (IoT) data derived from smartphones, wearables, and ambient sensors. Managing this vast amount of data, encompassing diverse data types and varying time scales, is crucial. Moreover, compliance with regulatory and contractual obligations is essential. The proposed data fabric comprises an architecture and a toolkit that facilitate the integration of heterogeneous data sources, across different environments, to provide a unified view of the data in dashboards. Furthermore, the data fabric supports the development of reusable and configurable data integration components, which can be shared as open-source or inner-source software. These components are used to generate data pipelines that can be deployed and scheduled to run either in the cloud or on-premises. Additionally, we present the implementation of our data fabric in a home-based telemonitoring research project involving older adults, conducted in collaboration with the University of California, San Diego (UCSD). The study showcases the streamlined integration of data collected from various IoT sensors and mobile applications to create a unified view of older adults' health for further analysis and research.
\end{abstract}



\section{Introduction}
The healthcare industry is experiencing great changes driven by the rise of digital health technologies and the increasing adoption of Internet of Things (IoT) devices. However, despite the potential for this combination of technologies to revolutionize healthcare by improving patient outcomes, there are still several key challenges related to data collection, organization, and accessibility, in addition to meeting regulatory compliance and contractual obligations.

To address these challenges, various groups have suggested a data fabric architecture-based approach for data life cycle management in healthcare scenarios. For example, Macias et al. developed a smart healthcare digital twin-based data fabric architecture that monitors the spread of COVID-19 virus in a nursing home \cite{Macias2022}. Ahouandjinou et al. proposed a hybrid architecture for a visual patient monitoring system in the Intensive Care Unit (ICU) to improve intensive health care \cite{Ahouandjinou2016}. Other frameworks are intelligent such as the data pipeline model which incorporates machine learning to automatically monitor, detect, mitigate, and alarm issues that arise at different stages of the data pipeline \cite{Raj2020}. However, while these proposed data fabric implementations are effective, they lack flexibility for users to choose between on-premises or cloud data integration. They also do not support the option to use different workflows engines, such as Kubeflow to manage machine learning (ML) and Machine Learning Operations (MLOps) pipelines, or Apache Airflow to manage complex data engineering pipelines at scale. Additionally, the data integration code cannot be easily reused, in a no-code / low-code approach, making the user training and usage of the data fabric tools and software for new projects more difficult.


In this paper, we present a data fabric solution that provides an end-to-end method to integrate heterogeneous data from EHRs, mobile applications, IoT sensors and wearable devices. We also showcase an example of the data fabric implementation in an older adult research study conducted in collaboration with UCSD with the purpose to deepen our understanding of the cognitive, social, and mobility impacts of clinical and subclinical sleep disturbance such as sleep apnea.

\section{Data Fabric for Digital Health}

 The data fabric for digital health provides a collection of tools and software that moves data through several stages: data collection, data storage, data integration, and data access with tight data governance. The architectural layers of the data fabric along with their corresponding components are shown in Figure 1. The data fabric can be deployed in different environments while sharing the same data access layer for insights in a self-service manner. 

\begin{figure}[htbp]
\centering
\includegraphics[width=0.9\linewidth]{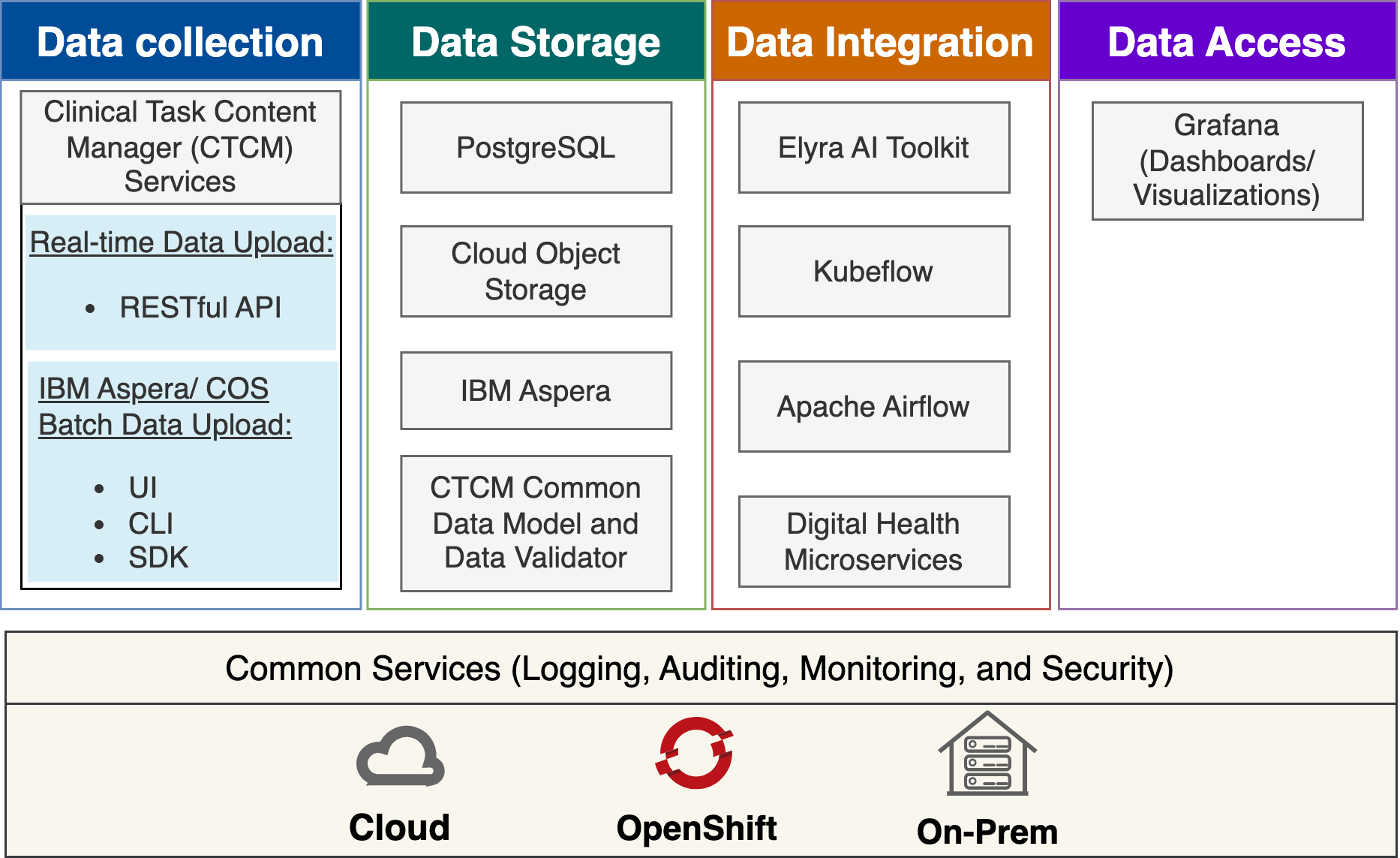}
\caption{Data Fabric Architecture. The data fabric makes use of available managed common services in the cloud or on-premises infrastructure, or common services deployed in an OpenShift\textsuperscript{\textregistered} container platform.}
\label{Figure5}
\end{figure}

Data collection is orchestrated using the clinical task content manager (CTCM) service which provides data governance and  supports real-time data upload via RESTful application programming interfaces (APIs) and batch data upload via a user interface (UI), command line interface (CLI) or software development kit (SDK). The CTCM validates input data against a set of configurable system-defined Critical Input Data Elements (CIDE) from the CTCM's common data model. CTCM also mandates Critical Output Data Elements (CODE) for the pipelines' data output, which includes a metadata file for data discovery purposes. These elements do not contain sensitive information, PHI/PII, and are used for analytics and reporting purposes via dashboards. The CODE comply with a digital health vocabulary used across the environment and enforced across services. The vocabulary is built using a crowdsourcing approach as new data elements are introduced in clinical study pipelines. 

The data storage uses an IBM cloud object storage (COS), a s3-based object storage bucket that is integrated with IBM Aspera\textsuperscript{\textregistered} to provide high speed data transfer. The datastore decouples the metadata from the data, by linking relational database records with IBM COS objects. The metadata elements are stored in a PostgreSQL relational database, while the data, unstructured or structured, such as audio, video, or sensor text data are stored in their native format in an IBM COS bucket. 

The data integration layer uses the Elyra AI Toolkit to facilitate the creation and deployment of data pipelines, for data integration and analytics, that can run locally or at scale using Kubeflow or Apache Airflow. Kubeflow and Apache Airflow enable the execution, management, and monitoring of pipelines or workflows at scale using a container platform that can run in the cloud or on-premises. Elyra pipeline nodes can invoke internal or external services via APIs or SDKs such as IBM Cloud Watson speech to text service or a Ray cluster used for at scale parallel processing. Elyra AI Toolkit enables the re-usability or modification of pipeline nodes, written in Python or R programming language, thereby facilitating the development of new data pipelines and leveraging the advanced analytics and medical knowledge from previous digital health use cases in a no-code / low-code approach. Elyra pipelines can be executed via a GUI in JupyterLab or CLI. Elyra AI Toolkit is integrated with Git, for full pipeline and pipeline nodes version control.

The data access layer uses Grafana to facilitate the delivery and analysis of data insights, via dashboards, across the environments using the data fabric. Grafana dashboards are configured to visualize data based on CTCM's configurable system-defined CODE. Each environment, using the data fabric, defines an outbound IBM COS bucket from where Grafana can read the data. A hybrid cloud approach can be implemented where the data fabric is deployed in a cloud or on-premises infrastructure that is close to where the data is collected, stored and integrated at the edge, as depicted in Figure 2. Additionally, Keycloak, an Identity Provider (IdP) that offers user Identity and Access Management (IAM) capabilities at scale, is integrated with Grafana.  

\begin{figure}[htbp]
\centering
\includegraphics[width=0.9\linewidth]{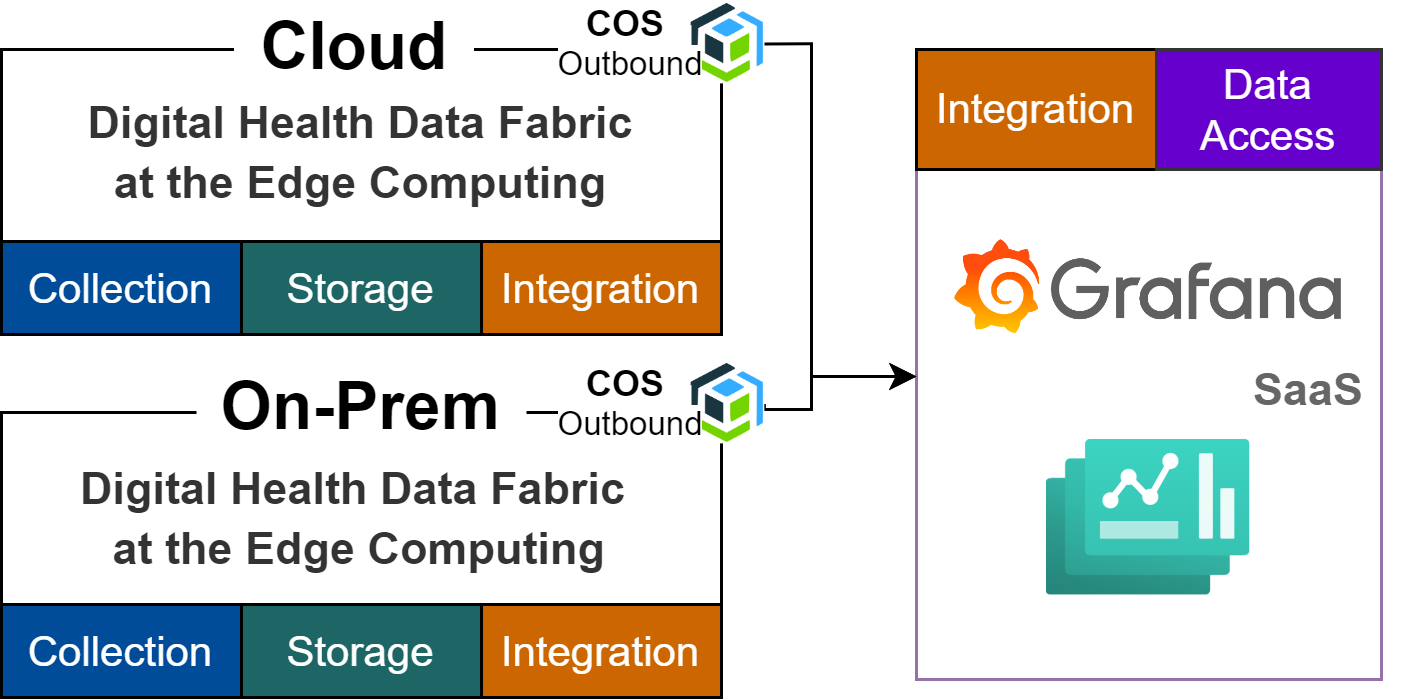}
\caption{Depiction of cloud and on-premises data fabric with distributed and composable components.}
\label{Figure5}
\end{figure}


In the state of the art, IBM Cloud Pak\textsuperscript{\textregistered} for Data (CP4D) provides a scalable enterprise grade data fabric architecture and offers IBM products with emphasis in data access and hybrid cloud. In the other hand, the digital health data fabric emphasizes security, trust and governance. It can minimize the complexity of the components needed to deploy and to manage for security and privacy controls, with flexibility to run in-place in a localhost or in standalone containers. This flexibility enables the data fabric tools to be deployed in run time environments managed by third party vendors or internal privacy and security teams within an organization to facilitate self-service data sharing for remote health monitoring. Table 1 lists the building blocks and functionality by the digital health data fabric implementation, and the IBM data fabric \cite{ibmdatafabric}. Data collection, which is critical for our IoT data driven digital health solution, is not likely to be added to CP4D. However, we acknowledge the need for, and plan to add, data catalog and data lineage support to our fabric. A catalog of data products along with mechanisms to facilitate access will provide the groundwork for data virtualization.

\begin{table}[tbp]
\centering
\caption{A functionality comparison between IBM data fabric and the data
fabric described in this work}
\includegraphics[width=1.0\linewidth]{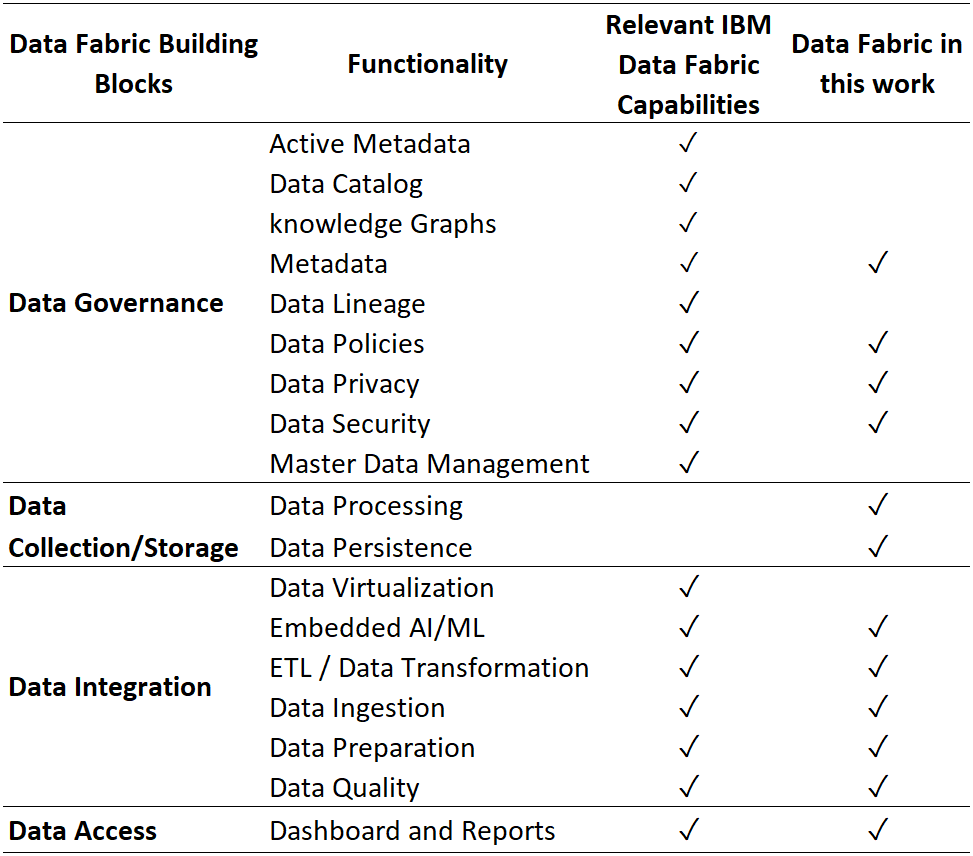}
\label{Table1}
\end{table}

\section{Implementation of data fabric in older adult home-based telemonitoring study}

\begin{figure}[htbp]
\centering
\includegraphics[width=.98\linewidth]{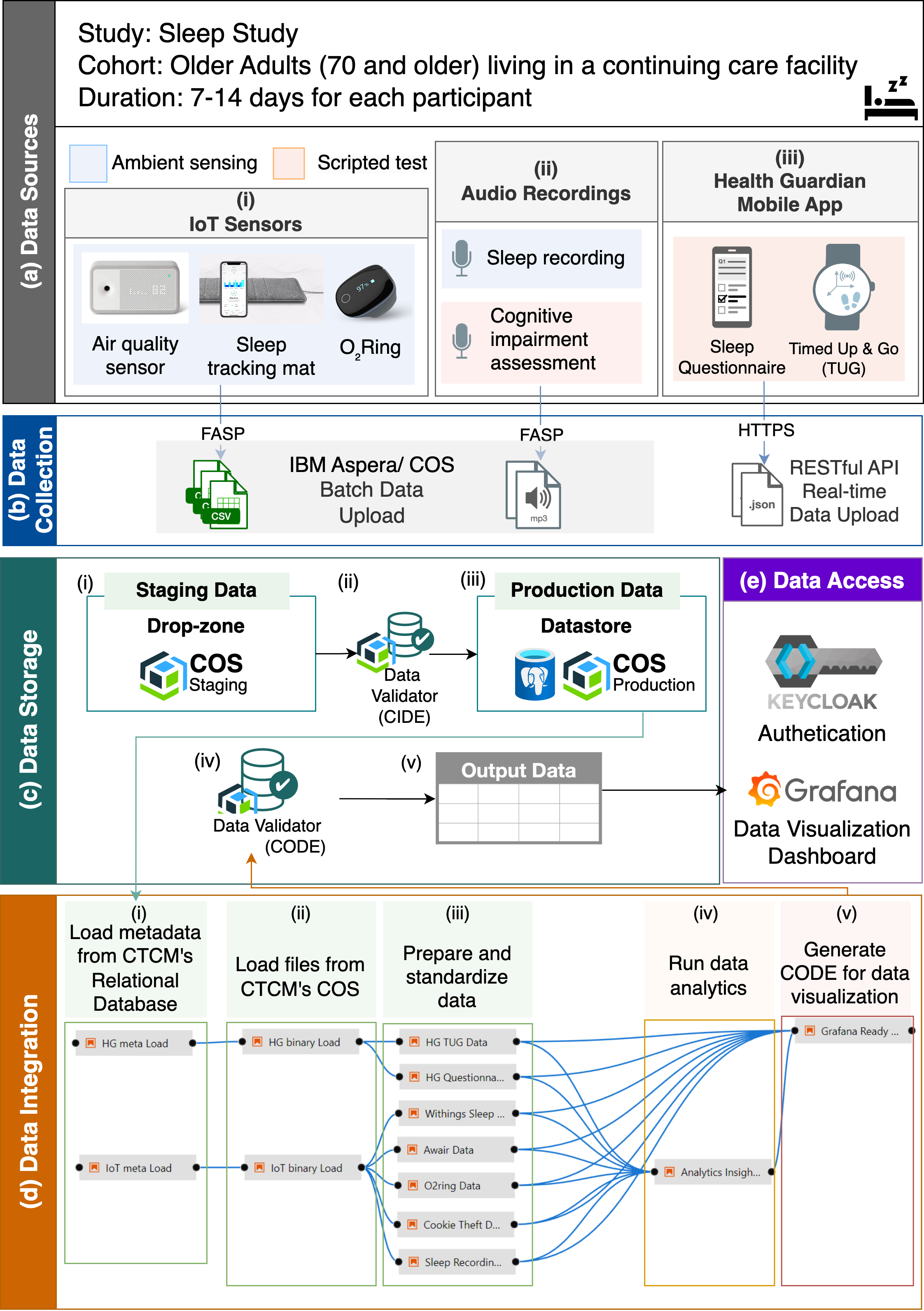}
\caption{(a) An overview of a home-based telemonitoring study. (b) The data collection process. (c) A data storage solution that tailors for storing multiple sources and formats of data. (d) Data integration pipeline for (i-iii) data preparation, (iv) data analytics and (v) generating CODE. (e) Data access using Grafana as a dashboard engine.}
\label{Figure5}
\end{figure}

An implementation of the data fabric discussed in this paper is a home-based telemonitoring study conducted in collaboration with UCSD and the IBM Digital Health team. The sensors used in this home-based telemonitoring study can be classified as ambient sensors or scripted tasks that require the participant's involvement \cite{Wen2022} (Figure 3a). Data ingestion was done either through a batch or real-time mode of operation, depending on the data source (Figure 3b). The data was stored in staging and production datastores, validated by CIDE (Figure 3c). Finally, the data were extracted, validated against CODE (Figure 3d), and visualized in a Grafana dashboard (Figure 3e).

\section{Conclusions}
In conclusion, we have presented a data fabric architecture with supporting tools that addresses many of the challenges that arise in the adoption of digital health technologies and IoT devices for research and clinical purposes. We demonstrated the effectiveness of our data fabric solution in an older adult home-based telemonitoring research study.

\section*{Acknowledgment}

The authors would like to thank Bo Wen from IBM for his insights and discussion on clinical task management engineering and acknowledge support from the IBM Cognitive Horizons Network and the IBM Research Accelerated Discovery Department.

\bibliography{DataFrabric2023-cameraReady_v1}
\bibliographystyle{IEEEtran}

\end{document}